\documentclass[11pt]{article}
\usepackage{graphicx}

\begin{document}

\def\xslash#1{{\rlap{$#1$}/}}
\def \p {\partial}
\def \dd {\psi_{u\bar dg}}
\def \ddp {\psi_{u\bar dgg}}
\def \pq {\psi_{u\bar d\bar uu}}
\def \jpsi {J/\psi}
\def \psip {\psi^\prime}
\def \to {\rightarrow}
\def \lrto{\leftrightarrow} 
\def\bfsig{\mbox{\boldmath$\sigma$}}
\def\DT{\mbox{\boldmath$\Delta_T $}}
\def\xit{\mbox{\boldmath$\xi_\perp $}}
\def \jpsi {J/\psi}
\def\bfej{\mbox{\boldmath$\varepsilon$}}
\def \t {\tilde}
\def\epn {\varepsilon}
\def \up {\uparrow}
\def \dn {\downarrow}
\def \da {\dagger}
\def \pn3 {\phi_{u\bar d g}}

\def \p4n {\phi_{u\bar d gg}}

\def \bx {\bar x}
\def \by {\bar y}



\begin{center} 
{\Large\bf   On the singular behavior of the chirality-odd twist-3 parton distribution $e(x)$  }
\par\vskip20pt
 J.P. Ma$^{1,2,3}$ and G.P. Zhang$^{4}$    \\
{\small {\it
$^1$ CAS Key Laboratory of Theoretical Physics, Institute of Theoretical Physics, Chinese Academy of Sciences, P.O. Box 2735, 100190 Beijing, China\\
$^2$ School of Physical Sciences, University of the Chinese Academy of Sciences, 100049 Beijing, China\\
$^3$ School of Physics and Center for High-Energy Physics, Peking University, 100871 Beijing, China\\
$^4$ Department of Physics, Yunnan University, Kunming, 650091 Yunnan, China}} \\
\end{center}
\vskip 1cm
\begin{abstract}
The first moment of the chirality-odd twist-3 parton distribution  $e(x)$ is related to the pion-nucleon $\sigma$-term, which is important for phenomenology.  However, the possible existence of a singular contribution proportional to $\delta(x)$ in the distribution prevents the determination of the $\sigma$-term with $e(x)$ extracted from  experimental data. 
There are two approaches to show the existence: the first one is based on an operator identity; the second  one is based 
on a perturbative calculation of a single quark state with finite quark mass.  We show that all contributions  proportional to $\delta (x)$ in the first approach are canceled. For the second approach we find that $e(x)$ of a multiparton state with a massless quark has no contribution with $\delta (x)$.  Considering that a proton is essentially  a multiparton state,  the effect of the contribution with $\delta(x)$ is expected to be  suppressed by light quark masses with arguments 
from perturbation theory.  A  detailed discussion of the difference between  cut diagrams and uncut diagrams
of $e(x)$  is provided.

 \end{abstract}      
\vskip 5mm
\noindent

\par\vskip20pt
\noindent
{\bf 1. Introduction} 
\par\vskip5pt
Cross sections of high-energy scattering involving hadrons can be predicted by use of the quantum chromodynamics (QCD) factorization theorem. 
At the leading power,  they can be predicted in the form of convolutions of perturbative coefficient functions with twist-2 parton distributions. In numerous experiments, these twist-2 parton distributions have been well studied and provide important information about the inner structure of hadrons.  At the next-to-leading power, twist-3 parton distributions are involved. These distributions contain more information than do twist-2 parton distributions, but are little known from experiments.

\par 
In this letter we study the chirality-odd twist-3  parton distribution $e(x)$. 
The most interesting quantity related to $e(x)$ is the pion-nucleon $\sigma$-term, determined by the first moment  of $e(x)$. This quantity gives important information about explicit chiral symmetry breaking of QCD \cite{JaJi}. It is also  phenomenologically important for searching for physics beyond the Standard Model.  
With the experimentally determined distribution $e(x)$, one can determine the $\sigma$-term in principle. 
But it seems impossible because $e(x)$ can have a contribution proportional to $\delta(x)$. 
Such a contribution around $x\sim 0$ cannot be determined experimentally.  Therefore, it is not possible 
to determine the $\sigma$-term from $e(x)$ extracted from experimental data. 
 There are two different approaches 
in the literature showing that $e(x)$ has a contribution proportional to $\delta(x)$.  
Because of its importance, we examine here  the existence of the $\delta(x)$-contribution.

\par 
The effects of higher-twist parton distributions are, in general, suppressed in high-energy scattering. Therefore, it is expected that 
their determination  is difficult. 
However,  with  experimental progress, 
there is already some information about $e(x)$ from experiments. The first extraction is given in \cite{EGS} 
from semi-inclusive deeply inelastic scattering    experiments with the CEBAF Large Acceptance Spectrometer (CLAS) \cite{CLAS1}.  The distribution 
has also been determined from dihadron production studied with in a CLAS experiment \cite{CO}.   With high-luminosity facilities such as as those at the Thomas Jefferson National Accelerator Facility \cite{JLab} and the planned Eelectron-Ion Collider (EIC) \cite{Eic} and the Electron-Ion Collider in China (EicC) \cite{EicC},   
twist-3 parton distributions can be studied more precisely. 
\par 
Our work is organized as  follows: In Sect.~2 we give definitions of chirality-odd twist-3 parton distributions and relations between these distributions. We show that the relation between $e(x)$ and other parton distributions does not explicitly have a contribution proportional to $\delta (x)$. In Sect.~3 we study $e(x)$ of a single quark state. It is shown that the results for $e(x)$ calculated from cut and uncut diagrams  should be the same.  In Sect.~4 we present our results for a multiparton state, which does not have any contribution proportional to $\delta (x)$. This indicates that the contribution with $\delta(x)$ is expected to be  suppressed by light quark masses in the case of a real hadron.  A brief summary of our work is given in Sect.~5.     

\par\vskip5pt 
\noindent 
{\bf 2. Definitions and operator relations} 
\par\vskip5pt  
We consider  a proton moving fast in the $z$-direction.  
We use the  light-cone coordinate system, in which a
vector $a^\mu$ is expressed as $a^\mu = (a^+, a^-, \vec a_\perp) =
((a^0+a^3)/\sqrt{2}, (a^0-a^3)/\sqrt{2}, a^1, a^2)$ and $a_\perp^2
=(a^1)^2+(a^2)^2$. The transverse metric is given by  
  $g_\perp^{\mu\nu} = g^{\mu\nu} - n^\mu l^\nu - n^\nu l^\mu$, where the two vectors are defined as
  $l^\mu =(1,0,0,0)$ and $n^\mu =(0,1,0,0)$.  In this coordinate system, the momentum of  the proton is given by $P^\mu =(P^+,P^-,0,0)$. 
We introduce the gauge link:
\begin{equation} 
{\mathcal L}_n (x) = P \exp \biggr \{-i g_s \int_0^\infty d\lambda n\cdot G( \lambda n + x) \biggr \}. 
\end{equation} 
With the gauge link, one can define the distributions in a gauge-invariant way. 
There are four chirality-odd distributions at twist-3 for an unpolarized proton, which are defined as   
\begin{eqnarray} 
 M e(x)   &=&  P^+ \int\frac{d \lambda }{4\pi} e^{ i\lambda x P^+ } 
   \langle P \vert \bar \psi (0 )  {\mathcal  L_n}^\dagger (0 )   {\mathcal  L_n} (\lambda n ) \psi (\lambda n ) \vert P\rangle,
 \nonumber\\  
\tilde T_F  (x_1,x_2) &=&   g_s\int\frac{d \lambda_1 d\lambda_2}{4 \pi} e^{-i \lambda_1 x_1 P^+ -i \lambda_2 (x_2-x_1) P^+} 
\nonumber\\ 
   && \langle P\vert \bar \psi (\lambda_1  n) {\mathcal  L_n}^\dagger (\lambda_1  n )  {\mathcal  L_n} (\lambda_2 n ) \left ( i \gamma_{\perp \mu}\gamma^+ \right )   G^{+\mu}(\lambda_2  n)  {\mathcal  L_n}^\dagger  (\lambda_2 n )    {\mathcal  L_n} (0  )  \psi (0) \vert P\rangle, 
 \nonumber\\
  E_D (x_1,x_2)  &=&   P^+\int\frac{d \lambda_1 d\lambda_2}{4 \pi} e^{-i \lambda_1 x_1 P^+ -i \lambda_2 (x_2-x_1) P^+} 
\nonumber\\ 
   && \langle P\vert \bar \psi (\lambda_1  n) {\mathcal  L_n}^\dagger (\lambda_1  n )  {\mathcal  L_n} (\lambda_2 n ) \left ( i \gamma_{\perp \mu}\gamma^+ \right )   D_\perp^\mu (\lambda_2 n)   {\mathcal  L_n}^\dagger  (\lambda_2 n )    {\mathcal  L_n} (0  )  \psi (0) \vert P\rangle, 
 \nonumber\\
  E_\partial (x) &=&  P^+ \int\frac{d \lambda }{4\pi} e^{-i\lambda  x P^+ } 
   \langle P \vert \bar \psi ( \lambda n)  {\mathcal  L_n}^\dagger (\lambda n )   i \gamma_{\perp \mu}\gamma^+ \partial^\mu {\mathcal  L_n} (0) \psi (0) \vert P\rangle,
 \label{DEFT3}   
 \end{eqnarray}  
with the covariant derivative
\begin{equation}
    D^\mu (x) = \partial^\mu + i g_s G^\mu (x).    
\end{equation}    
From symmetries, one can derive
\begin{equation} 
  \tilde T_F (x_1,x_2) = \tilde T_F (x_2,x_1), \quad E_D (x_1,x_2) = - E_D (x_2, x_1). 
\end{equation} 
The defined four parton distributions are not independent. One can derive the relation
\begin{eqnarray}
  E_D (x_1,x_2) =  -i  \pi \delta (x_1-x_2)  E_\partial (x_1)  + \frac{1}{ x_2-x_1 + i\varepsilon} \tilde T_F(x_1,x_2). 
\end{eqnarray}  
Taking the principal value of the distribution $1/(  x_2-x_1 + i\varepsilon)$, one has           
\begin{equation} 
  E_D (x_1,x_2) = \tilde T_F(x_1,x_2) P \frac{1}{x_2-x_1}, \quad \tilde T_F (x,x) = E_\partial (x). 
\end{equation}   
Therefore, only one of the last three twist-3 distributions in Eq.~(\ref{DEFT3}) is independent. 
The defined distributions depend on the renormalization scale $\mu$. The dependence 
was studied in  \cite{BBKT,KoNi,BD,Beli1,MWZ}.

\par 
Now we focus on the distribution $e(x)$.  The factor $M$ in the definition of $e(x)$ is a scale 
factor to make $e(x)$ dimensionless.  It is convenient to take $M$ as the proton mass. In principle it 
can be any mass quantity at the order of $\Lambda_{QCD}$.  It is easy to find the first moment:
\begin{equation} 
  \int_{-1} ^1 d x e(x) = \frac{1}{2 M} \langle P \vert \bar \psi \psi \vert P \rangle. 
 \label{SRE}  
\end{equation} 
By taking $M$ as the proton mass and summing different flavors of light quarks, we find the first moment is related to the pion-nucleon $\sigma$-term, which is important for phenomenology.  The $\sigma$-term at the moment cannot be directly accessed experimentally. This term can only be extracted from pion-nucleon scattering \cite{EXS} or calculated with lattice QCD (e.g., as shown in \cite{EXSLA}).  
The sum rule in Eq.~(\ref{SRE}) gives the  possibility to determine the $\sigma$-term by using $e(x)$ extracted from experiments.  However, this possibility may not exist. 
There is evidence that $e(x)$ contains a contribution proportional to $\delta (x)$.  If this is the case, then one can never determine the integral in the sum rule by experiments and hence the $\sigma$-term,  because the region with $x\sim 0$ cannot be accessed experimentally.  In this case, the sum rule is violated.    

\par 
As already mentioned, there are two approaches to show that $e(x)$ contains a contribution proportional to 
$\delta(x)$.  One is given in \cite{ex}. To see such a contribution, one starts with the identity for the operator in the definition of $e(x)$: 
\begin{equation} 
\bar \psi (0 )  {\mathcal L_n}^\dagger (0 )   {\mathcal  L_n} (\lambda n ) \psi(\lambda n) 
   =\bar \psi(0) \psi(0) + \int_0^\lambda d \sigma  \frac{ \partial}{\partial \sigma } \biggr ( \bar \psi ( 0 )  {\mathcal L_n}^\dagger ( 0 )   {\mathcal  L_n} ( \sigma n ) \psi( \sigma  n) \biggr ).
\label{IDI}    
 \end{equation}     
If we take the matrix element of Eq.~(\ref{IDI}), it is found that  the integral can be expressed with a twist-3 quark-gluon parton distribution and a twist-2 quark distribution \cite{KoTa,BaBr,BrFi}.  From this identity, one may conclude that $e(x)$ has a contribution proportional to $\delta (x)$, which is given by the first term in Eq.~(\ref{IDI}) \cite{ex}:  
\begin{equation} 
M e(x) = \frac{1}{2} \delta (x) \langle P \vert \bar \psi (0) \psi (0) \vert P\rangle + \cdots,  
\label{SIN} 
\end{equation} 
where $\cdots$ denotes the contribution of the twist-3 quark-gluon operator and that of the twist-2 quark distribution. 
It is clear that this conclusion is correct only if the remaining contributions in Eq.~(\ref{IDI}) contain no term proportional to 
$\delta(x)$.  However,  there are terms with $\delta(x)$ in the remaining contributions. 
If these terms are canceled, then there is no contribution proportional to $\delta(x)$. 
\par 
We need to carefully examine the remaining contribution.  With a little algebra the derivative of the operator in Eq.~(\ref{IDI}) can be written as
\begin{eqnarray} 
\frac{ \partial}{\partial \sigma } \biggr ( \bar \psi ( 0 )  {\mathcal L_n}^\dagger ( 0 )   {\mathcal  L_n} ( \sigma n ) \psi( \sigma n) \biggr )
  &=& \frac{1}{2} \bar \psi ( 0 )  {\mathcal L_n}^\dagger ( 0 ) {\mathcal  L_n} ( \sigma n )   \biggr (  \gamma_\perp^\nu \gamma^+ D_\nu  + \gamma^+ \gamma\cdot D \biggr ) \psi( \sigma n)
\nonumber\\
   && - \frac{1}{2} \partial^+ \biggr ( \bar \psi(0) {\mathcal L_n}^\dagger ( 0 ) \biggr ) \gamma^- \gamma^+    {\mathcal  L_n} ( \sigma n ) \psi( \sigma  n)  
\nonumber\\
   && +  \frac{1}{2} \partial^+ \biggr ( \bar \psi(0) {\mathcal L_n}^\dagger ( 0 ) \gamma^- \gamma^+   {\mathcal  L_n} ( \sigma  n ) \psi( \sigma  n) \biggr ). 
\end{eqnarray}  
The term in the last line is a total derivative term. It gives no contribution when sandwiched into a state.  
 The derivative in the second line multiplied by $\gamma^-$ can be expressed with the Equation Of Motion(EOM)  as
 \begin{equation}
    \partial^+ \biggr ( \bar \psi(0) {\mathcal L_n}^\dagger ( 0 ) \biggr ) \gamma^- =- ( \overline { D_\nu  \psi(0)} )  {\mathcal L_n}^\dagger ( 0 )  \gamma_\perp^\nu + i m_q  \bar \psi (0) {\mathcal L_n}^\dagger ( 0 ),\end{equation}
where $m_q$ is the quark mass. 
Using the identity 
\begin{equation}
 \int \frac{d \lambda_2 d y }{2\pi}  e^{ - i ( \lambda_2 -\lambda )  y } {\mathcal L}_n^\dagger ( \lambda n) {\mathcal L}_n (\lambda_2 n) =1,  
\label{FDI}  
\end{equation}  
we find the matrix element of the derivative term in Eq.~(\ref{IDI}) becomes
\begin{eqnarray} 
&&  \frac{ \partial}{\partial \sigma } \langle P \vert  \bar \psi ( 0 )  {\mathcal L_n}^\dagger ( 0 )   {\mathcal  L_n} ( \sigma n ) \psi( \sigma  n) \vert P\rangle  
\nonumber\\ 
  &&=  \frac{1}{2} \int \frac{ d y d \lambda_2 }{2\pi} e^{-i y(\lambda_2- \sigma) } \langle P \vert \bar \psi ( 0 )  {\mathcal L_n}^\dagger ( 0 )  \gamma_\perp^\nu \gamma^+ {\mathcal  L_n} ( \lambda_2  n )  D_\nu(\lambda_2 n) 
    {\mathcal L}_n ^\dagger ( \lambda_2 n ) {\mathcal L}_n (\sigma  n)    \psi( \sigma  n)  \vert P\rangle 
\nonumber\\
   && \quad  -  \frac{1}{2}  \int \frac{ d y d \lambda_2 }{2\pi} e^{-i y \lambda_2 }    \langle P \vert \bar   \psi(0) {\mathcal L_n}^\dagger ( 0 )  \gamma_\perp^\nu  \gamma^+  {\mathcal  L_n} ( \lambda_2 n ) D_{\nu} (\lambda_2 n)  {\mathcal  L_n}^\dagger ( \lambda_2 n )  {\mathcal  L_n} ( \sigma n ) \psi( \sigma  n) \vert P \rangle 
\nonumber\\
 &&\quad   - i  m_q   \langle P \vert\bar \psi (0) {\mathcal L_n}^\dagger ( 0 ) \gamma^+  {\mathcal  L_n} ( \sigma  n ) \psi(\sigma n)  \vert P \rangle.     
\end{eqnarray} 
The operator in the second and third lines is the operator used to define the twist-3 distribution 
$E_D(x_1,x_2)$ in Eq.~(\ref{DEFT3}).  The operator in the last line is the one used to define the twist-2 parton distribution $f_q$.   Finally, we can derive the following relation:
\begin{eqnarray}   
M e(x)  
  &=&   \delta (x) \biggr (  \frac{1}{2}  \langle P\vert \bar \psi (0) \psi(0) \vert P\rangle -\frac{1}{4\pi}  
  \int\frac{ d x_1 d x_2}{x_1 x_2} (x_2-x_1) E_D(x_1,x_2) -  m_q \int \frac{d x_1 }{x_1} f_q (x_1) \biggr ) 
\nonumber\\
   && + \frac{1}{4\pi} \int d x_1 d x_2 E_D (x_1,x_2) \biggr ( \frac{1}{x_1} \delta (x-x_1) -\frac{1}{x_2} 
   \delta (x -x_2) \biggr )
   + \frac{m_q}{ x} f_q(x). 
\label{ER14}        
\end{eqnarray}  
Therefore, there are three terms with $\delta(x)$, not only the one given in Eq.~(\ref{SIN}). 

\par 
Since there are three terms with $\delta(x)$, it is possible that their sum is zero so that $e(x)$ contains no contribution proportional to $\delta(x)$. One can show that the sum is zero. For this purpose we can write  the quark field  as the sum of the  plus-component and  the minus-component, which are defined as
\begin{equation} 
   \psi^{(+)}(x) = \frac{1}{2} \gamma^- \gamma^+ \psi(x), \quad   \psi^{(-)} (x) = \frac{1}{2} \gamma^+ \gamma^- \psi (x). 
\end{equation} 
With these components the matrix element of $\bar \psi \psi$ becomes
\begin{equation} 
 \langle P\vert \bar \psi (0)   \psi (0) \vert P\rangle =   \langle P\vert \bar \psi^{(+)}  (0)   \psi^{(-)}  (0)\vert P\rangle  + \langle P\vert \bar \psi^{ (-) }  (0)   \psi^{(+ )}  (0)\vert P\rangle. 
 \end{equation}  
 The two components are not independent.  
With use of  EOM,  the minus-component can be expressed with the plus-component combined with gauge fields: 
\begin{eqnarray} 
   \psi^{(-)}(x) = \frac{1}{2} {\mathcal L}_n^\dagger (x)  \int_0^{\infty} d \lambda \biggr [ {\mathcal L}_n \gamma^+ \biggr ( \gamma_\perp^\mu  D_\mu + i m_q \biggr )  \psi^{(+)} \biggr ] ( \lambda n +x).
\label{17}    
\end{eqnarray} 
In this solution we assume as usual that the minus-component of $\psi$ is zero at $x^- =\infty$. 
Before our summary, we will discuss  the case that $\psi^{(-)} (x) $ is nonzero at  $x^- =\infty$.        
Using this expression and the identity in Eq.~(\ref{FDI}), we can write the matrix element in the form
\begin{eqnarray}
 \langle P\vert \bar \psi (0)   \psi (0) \vert P\rangle &=&  \frac{1}{2} \int   \frac{d \lambda d \omega }{2\pi} e^{-i\omega \lambda } \frac{i}{\omega + i\varepsilon}  \int \frac{d \lambda_2 d y }{2\pi} d y e^{ - i ( \lambda_2 -\lambda )  y } 
\nonumber\\  
&&  \langle P \vert \bar \psi (0)  {\mathcal L}_n^\dagger (0)  {\mathcal L}_n  (\lambda n) \gamma^+\gamma_\perp^\mu  D_\mu (\lambda n)    {\mathcal L}_n^\dagger ( \lambda n) {\mathcal L}_n (\lambda_2 n)  \psi  ( \lambda_2 n )  \vert P\rangle 
\nonumber\\
  && + \frac{1}{2} i m_q \int_0^\infty d\lambda \langle P\vert \bar \psi (0) {\mathcal L}_n^\dagger (0) \gamma^+ {\mathcal L}_n (\lambda n) \psi (\lambda n) \vert P\rangle
 + \mbox{ h.c.}  
\end{eqnarray} 
The operator in the second and third lines is used to define $E_D$ and $f_q$, respectively.     
Therefore, the matrix element is related to $E_D$ and $f_q$. The relation is 
\begin{equation} 
\  \langle P\vert \bar \psi (0) \psi(0) \vert P\rangle = \frac{1}{2\pi}  
  \int\frac{ d x_1 d x_2}{x_1 x_2} (x_2-x_1) E_D(x_1,x_2) +  2 m_q \int \frac{d x_1 }{x_1} f_q (x_1). 
\end{equation} 
This shows that the sum of the three terms with $\delta(x)$ in Eq.~(\ref{ER14}) is zero. The correct 
relation for $e(x)$ instead of that in Eq.~(\ref{ER14}) is
\begin{eqnarray}   
M e(x)  
   = \frac{1}{2\pi x} \int  d x_2 E_D (x,x_2)   + \frac{m_q}{ x} f_q(x) 
\label{E(x)}        
\end{eqnarray}  
without $\delta(x)$-terms explicitly.  

\par 
In $e(x)$,  integration is done over the transverse momentum of the parton. One can define a transverse-momentum-dependent parton distribution $e(x, k_\perp)$ by undoing the integration.  The defined  distribution 
has a  relation similar to that of $e(x)$ in Eq.~(\ref{ER14}) shown in \cite{PaRo}, where there are three terms with 
$\delta(x)$ corresponding to those in Eq.~(\ref{ER14}).  One can use the equation of motion as done above to show that the sum of the  three terms with 
$\delta(x)$ is zero. Therefore,  there is also no term with $\delta (x)$.  
\par 
From our analysis we obtain the relation for $e(x)$ in Eq.~(\ref{E(x)}) instead of that in Eq.~(\ref{SIN}). The relation obtained does not have the singular contribution around $x=0$ as given in Eq.~(\ref{SIN}).   However,  this does not imply 
that $e(x)$ is regular around $x=0$ because in Eq.~(\ref{E(x)}) there is a factor $x$ in the denominator. 
From perturbation theory one finds a singular contribution, which we discuss in the next section.

\par\vskip5pt
\begin{figure}[hbt]
	\begin{center}
		\includegraphics[width=8cm]{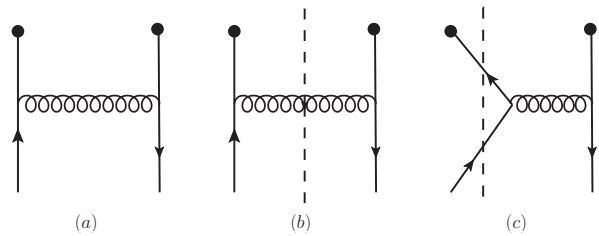}
	\end{center}
	\caption{(a): Uncut diagram for one-loop correction to $e(x)$ of a single quark state. (b) and (c):  Cut diagrams for one-loop correction to $e(x)$ of a single quark state. The black dots in all diagrams represent the insertion of the quark field in the definition of $e(x)$.  }
	\label{TreeL}
\end{figure}

\par 
\par\vskip5pt
\noindent 
{\bf 3. $e(x)$ of a single quark state} 
\par\vskip5pt 
In a study with perturbation theory in \cite{BuKo},  it was found that $e(x)$ of a single quark state 
has a $\delta(x)$-contribution.  Here, we examine this in detail.  It is straightforward to calculate the distribution of $e(x)$ of a quark state with momentum $p$ perturbatively.  Because $e(x)$ is a chirality-odd distribution, the quark must have a nonzero mass to obtain a nonzero result.  At tree level, the result is
\begin{equation} 
  M  e(x) = m_q \delta (1-x) + {\mathcal O}(\alpha_s). 
\label{21} 
\end{equation} 
At this order there is no singular contribution proportional to $\delta(x)$. 
At tree level we have the matrix element $\langle P\vert \bar\psi \psi \vert P\rangle =2 m_q$.  According 
to Eq.~(\ref{SIN}), we should have a $\delta (x)$-contribution at this order as $m_q \delta(x)$.  This indicates that 
the singular contribution, if it exists, cannot be that determined by the term in Eq.~(\ref{SIN}). 
 
\par 
We study the one-loop correction in the light-cone gauge $n\cdot G=0$, where  gauge links 
in Eq.~(\ref{DEFT3}) become unity. The contribution at one-loop level consists of two parts: one is the correction of external legs of the tree-level diagram; the other one is given by the diagram in Fig.~1(a). The contribution   
from Fig.~1(a) has a $\delta(x)$-term. The diagram is an uncut diagram. With a cut diagram one can miss the contribution with $\delta(x)$  \cite{AsBu}.  We will discuss the difference between calculations with cut diagrams and uncut diagrams.   
\par
The contribution from Fig.~1(a) is  
\begin{eqnarray} 
M e(x)\biggr\vert_{1a}  &=& g_s^2 C_F \frac{1}{2} \sum_s  \int \frac{d^4 q}{(2\pi)^4} \frac{1}{2} p^+ \delta( q^+ - xp^+) 
   \biggr \{  {\rm Tr}  \biggr [  \gamma_\alpha \frac{ \gamma\cdot q + m_q}{q^2-m_q^2 + i\varepsilon} \Gamma   \frac{ \gamma\cdot q + m_q}{q^2-m_q^2 + i\varepsilon} \gamma_\beta 
\nonumber\\   
   &&  u(p,s)\bar u(p,s) \biggr ]   \biggr (  g^{\alpha\beta} -\frac{ n^\alpha k^\beta }{n\cdot k} 
      - \frac{ n^\beta k^\alpha }{n\cdot k}  \biggr ) \biggr \}  \frac{-i}{ k^2 + i\varepsilon} ,    \quad q= p-k,  
\label{Fig1a}  
\end{eqnarray}
where $q$ is the momentum carried by the quark propagator connecting the black dots, and $k$ is the momentum carried by the gluon line.   For $e(x)$, $\Gamma$ is a unit matrix. 
We will keep 
only the leading order of $m_q$. The collinear divergence associated with the limit $m_q\to 0$ will be regularized 
with dimensional regularization.   
Working out the trace and the contraction of Lorentz indices,  we have
\begin{eqnarray} 
M e(x)\biggr\vert_{1a}  &=&  2m_q  g_s ^2 C_F p^+  \int \frac{d^4 q}{(2\pi)^4}  \delta( q^+ - xp^+)  \biggr (   \frac{-2p^+ } { k^+ } 
   \frac{ -i }{ (q^2-m_q^2 + i\varepsilon)  (k^2 + i\varepsilon )}    
   \nonumber\\
    && + \frac{-i}{(q^2-m_q^2 + i\varepsilon)^2}  + {\mathcal O}(m_q^2) \biggr ).  
\end{eqnarray} 
Analyzing the positions of poles in the complex $q^-$-plane, we easily find that the contribution from Fig.~1(a) 
is zero for $x>1$ and $x<0$ because  all poles are either in the upper half-plane or in the lower half-plane in these cases.    
The second term in in parentheses is proportional to the integral studied in detail in \cite{Yan}:
\begin{equation} 
 \int \frac{d q^-}{2\pi} \frac{-i }{ (  q^2 -m_q^2 + i\varepsilon)^2} =  \frac{1}{2} \delta (q^+) \frac{1} { q_\perp^2 + m_q^2},
 \end{equation} 
 with fixed $q^+= xp^+$.  This integral is zero for $x\neq 0$, but becomes singular at $x=0$. It is proportional to $\delta (x)$. This gives $e(x)$ a contribution proportional to $\delta(x)$. The calculation 
of the one-loop correction is straightforward. We obtain
\begin{eqnarray} 
  M e(x) &=&m_q \delta (1-x) + m_q \frac{\alpha_s C_F}{2\pi}\biggr (-\frac{2}{\epsilon_c} + \ln \frac{ \mu^2 e^\gamma}{4\pi \mu_c^2} \biggr )  \biggr [ \frac{1}{2} \delta (1-x) + \delta (x) + \theta(x) \theta(1-x) \frac{2 }{(1-x)_+} \biggr ] 
\nonumber\\  
    &&  + {\mathcal O} (\alpha_s^2 m_q ) + {\mathcal O} (\alpha_s m_q^3),
\label{25} 
\end{eqnarray}  
where the pole in $\epsilon_c =4-d$ represents collinear singularities.  As discussed before Eq.~(\ref{SRE}), 
one can take $M$ as the quark mass $m_q$ for the single quark state. Then $e(x)$ will depend on $m_q$ only through $\ln m_q$, which regularizes collinear singularities. Here, we have expanded the contribution in $m_q$. The collinear divergence is regularized with dimensional regularization. 
For convenience of our later discussion, we keep 
$M$ as an unspecified mass scale.  
The above result indicates that  $e(x)$ contains 
a contribution proportional to $\delta(x)$ from a perturbative calculation with a single quark state. It is also noted in \cite{BuKo} that without the contribution 
the sum rule in Eq.~(\ref{SRE}) cannot be satisfied.  

\par 
It is interesting to calculate the real correction of $e(x)$ with cut diagrams.  In such a calculation one usually  calculates only 
the cut diagram as shown in Fig.~1(b); that is, there is a cut cutting the gluon line. The cut implies that 
the contribution is obtained by replacement of $1/(k^2+i\varepsilon)$ with $-2\pi i \delta (k^2)$ in Eq.~(\ref{Fig1a}). 
It is straightforward to obtain the result:
\begin{eqnarray} 
  M e(x)\biggr\vert_{1b}  =  m_q \frac{\alpha_s C_F}{2\pi} \theta (1-x) \biggr (-\frac{2}{\epsilon_c} + \ln \frac{ \mu^2 e^\gamma}{4\pi \mu_c^2} \biggr )  \biggr ( 1 + \frac{1+x}{1-x} \biggr ). 
\label{1b}   
\end{eqnarray} 
This contribution has no $\delta(x)$-term as shown in \cite{BuKo}. 
 In the calculation of the uncut diagram in  Fig.~1(a),  one always finds 
that its contribution is zero for $x<0$ and $x>1$.   For Fig.~1(b), we have $k^+ >0$ because of the cut. 
This gives only the constraint that the contribution is zero for $x>1$.  It  does not give the constraint that the contribution is  zero for $x<0$. Therefore, the contribution is nonzero for $x<0$ as indicated by Eq.~(\ref{1b}). 
However, at this order of $\alpha_s$, it is expected that $e(x)=0$ for $x<0$. 
If we use the one-loop diagram in Fig.~1(b) for the factorization of deep inelastic scattering, the contribution from Fig.~1(b) with $x>0$ corresponds to the physical process $\gamma^* + q\to q$, where the initial quark 
carries the momentum fraction $x$.  The contribution from Fig.~1(b) with $x<0$ corresponds to the process $\gamma^* \to q \bar q$, where the antiquark  
carries the momentum fraction $x$. 
This is not allowed 
because the virtuality of the photon is negative. Therefore, $e(x)$ at the order considered here must be zero for $x<0$. The contribution from Fig.~1(b) for $x<0$ cannot be the 
only one. There is another cut diagram given in Fig.~1(c).  Beyond the order considered,
$e(x)$ can be nonzero for $x<0$.  

\par 
The contribution from Fig.~1(c) is obtained from the contribution from Fig.~1(a)  by replacing one of the $1/(q^2-m_q^2+i\varepsilon)$ expressions in Eq.~(\ref{Fig1a}) with $-2\pi i \delta (q^2-m_q^2)$.  Because of the cut, the contribution is nonzero only for $x<0$.  However, 
 it is difficult to calculate the contribution because one will encounter an undetermined factor such as  $\delta (q^2-m_q^2)/(q^2-m_q^2)$. This results in a divergent integral.    
We divide the contribution from Fig.~1(c) or any diagram into two parts:
 \begin{equation} 
    e (x) = e_g (x) + e_n (x),  
\end{equation}      
where $e_g (x) $ is the contribution with $g^{\alpha\beta}$ in Eq.~(\ref{Fig1a}) and $e_n (x)$ is the remaining contribution.  $e_g(x)$ is the contribution from Fig.~1(c) in Feynman gauge and has the  difficulty mentioned. In the calculation of $e_n(x)$ there is no such difficulty. It can be calculated in a straightforward way. The contribution from Fig.~1(c) and its complex conjugate diagram is
\begin{equation} 
M e_n (x) \biggr\vert_{1c + c.c }    =   - m_q\frac{\alpha_s}{2\pi}  C_F \theta (-x)  \biggr (-\frac{2}{\epsilon_c} + \ln \frac{ \mu^2 e^\gamma}{4\pi \mu_c^2} \biggr )  \frac{1+x}{1-x} . 
\end{equation} 
The contribution to $e_g$ from Fig.~1(c) and its complex conjugate diagram
can  be determined only up to an ill-defined integral:
\begin{equation} 
M e_g (x) \biggr\vert_{1c +cc}    =  - m_q C_F \frac{\alpha_s}{2\pi} \theta (-x)  \biggr [  \biggr (-\frac{2}{\epsilon_c} + \ln \frac{ \mu^2 e^\gamma}{4\pi \mu_c^2} \biggr )    + [ I_u (x) ]   \biggr ] 
      + {\mathcal O} (m_q^3), 
\label{1c}        
\end{equation} 
where $I_u(x)$ is the ill-defined integral
\begin{equation} 
I_u (x) =  (4\pi)^2 p^+  \int \frac{ dq^- d ^2 q_\perp}{ (2\pi)^3} \frac{\delta (q^2)}{q^2}  
  \frac{ 2 k^2 + q^2} { k^2},  
\end{equation}   
where  $k^+ $ is fixed as $(1-x) p^+$.  The result in Eq.~(\ref{1c}) has an ambiguity because  the integral has a term proportional to $\delta(q^2)/q^2$. 
We attempted to fix the ambiguity by regularizing the integral in different ways but without success.
However, this ambiguity 
can be fixed partly by the fact that $e(x)$ is zero at the order considered for $x<0$.  Summing all contributions, we have the one-loop real part: 
\begin{eqnarray} 
    Me (x) \biggr \vert_R &=& M e(x)\biggr\vert_{1b} + M e_n (x) \biggr\vert_{1c +cc} + M e_g (x) \biggr\vert_{1c +cc} 
\nonumber\\
   &=& m_q \frac{\alpha_s C_F}{2\pi} \theta (1-x) \theta (x)  \biggr (-\frac{2}{\epsilon_c} + \ln \frac{ \mu^2 e^\gamma}{4\pi \mu_c^2} \biggr )  \biggr ( 1 + \frac{1+x}{1-x} \biggr )
   - m_q C_F \frac{\alpha_s}{2\pi} \theta (-x)  [ I_u (x) ]. 
 \label{33}        
 \end{eqnarray}       
 The fact that  $e(x)$ is zero  for $x<0$ at the  order considered implies the function $I_u(x)$ must be  zero for $x\neq 0$. However, it can be nonzero at $x=0$. If the results from cut and uncut diagrams are the same, then $I_u (x)$ can be fixed and is proportional to $\delta (x)$. Below we show that this is indeed the case. 

\par 
\begin{figure}[hbt]
	\begin{center}
		\includegraphics[width=10cm]{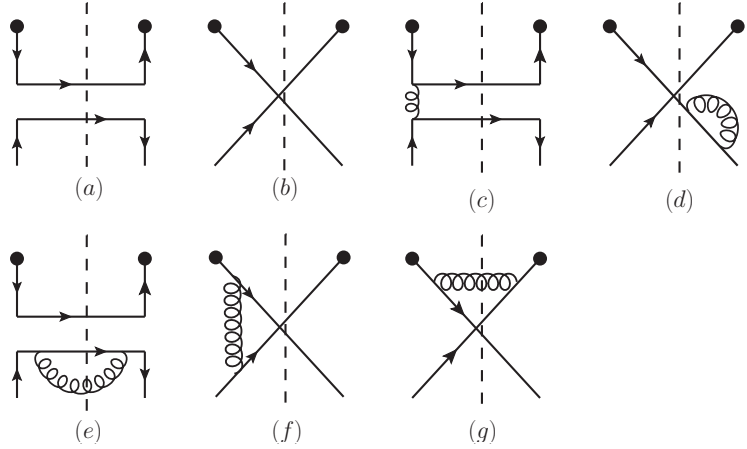}
	\end{center}
	\caption{Feynman diagrams for $\tilde e(x)$. The first two are at tree level.  The other diagrams are  for one-loop correction. In each diagram, the black dots denote the insertion of the quark field. The left black dot is for $\bar \psi$, while in Fig.~1 it is for $\psi$.     }
	\label{BE}
\end{figure}

 \par 
If one uses uncut diagrams to calculate $e(x)$, this implies that one uses the $T$-ordered product of operators to define  $e(x)$. In the light-cone gauge, the product is $T ( \bar \psi(\lambda n) \psi (0) ) $.  By use of cut diagrams, this implies that the product is not $T$-ordered. It is simply the product  $\bar \psi(\lambda n) \psi (0)$. 
The ordering along the time direction is the same as the ordering along the direction $n$ in our case. The difference between $e(x)$ defined with $T ( \bar \psi(\lambda n) \psi (0) )$ and $e(x)$ defined with $\bar \psi(\lambda n) \psi (0) ) $ can be given in the light-cone gauge as
\begin{eqnarray} 
 M \Delta  e(x)   &=&   P^+ \int\frac{d \lambda }{4\pi} e^{-i\lambda x P^+ } 
   \langle P \vert  T \biggr ( \bar \psi ( \lambda n)  \psi (0) \biggr )  - \bar \psi ( \lambda n)  \psi (0) \vert P\rangle,  
\nonumber\\
   &=&  - P^+  \int\frac{d \lambda }{4\pi} e^{-i\lambda x P^+ } \theta (-\lambda ) 
  \langle P \vert  \bar \psi ( \lambda n)   \psi (0)   + 
       {\rm Tr} \biggr [  \psi (0)  \bar \psi ( \lambda n)  \biggr ] \vert P\rangle.    
\end{eqnarray}       
 The difference is determined by the matrix element of the anticommutator of the two quark fields.   
 We introduce a new distribution $\tilde e(x)$ defined as
 \begin{equation}
M \tilde e (x) =P^+   \int\frac{d \lambda }{4\pi} e^{-i\lambda x  P^+ }
        \langle P \vert         {\rm Tr} \biggr [  \psi (0)   \bar \psi ( \lambda n)  \biggr ] \vert P\rangle.  
\end{equation}    
With this distribution the difference is expressed as
\begin{equation} 
  M \Delta e (x) = -\int dy  \frac{ i}{ y + i\varepsilon} \biggr ( e(x-y) + \tilde e(x-y) \biggr ). 
\end{equation}    
If we calculate $\tilde e(x)$ of a single quark as we did for $e(x)$  above, the contribution at  tree level is given by the first two diagrams in Fig.~2, where the intermediate state is a two-quark state.  Fig.~2(a) shows a disconnected diagram, which should be extracted or excluded. We have at tree level only the contribution from Fig.~2(b): 
\begin{equation} 
  M \tilde e(x) = - m_q \delta(1-x) + {\mathcal O}(\alpha_s). 
\end{equation}    
Compared with $e(x)$ in Eq.~(\ref{21}), there is an extra minus sign because of the order of the two-quark state, or because the two quark lines are crossed. Hence, at tree level, the difference $\Delta e (x)$ is zero.  
\par 
At one-loop level, there are disconnected diagrams such as that in Fig.~2(e). Their contributions should be excluded. 
The virtual corrections are from Fig.~2(c), (d), and (f).  Fig.~2(d)  represents the correction of the external lines,  
whose contribution is the same as the virtual correction to $e(x)$. The contribution from Fig.~2(c) is zero, because  a gluon cannot be coupled with a scalar operator.  The real correction is from Fig.~2(g). It can be calculated directly:  
\begin{eqnarray} 
  M \tilde e(x) \biggr\vert_{\ref{BE}g}   =   m_q \theta (x>1) \frac{\alpha_s C_F}{2\pi}\biggr (-\frac{2}{\epsilon_c} + \ln \frac{ \mu^2 e^\gamma}{4\pi \mu_c^2} \biggr )  \biggr ( 1  + \frac{1+ x }{1-x} \biggr )
 + {\mathcal O} (\alpha_s m_q^3). 
\end{eqnarray}   
The contribution from Fig.\ref{BE}(f)  has the discussed ambiguity in the case of $e(x)$. It can be expressed 
with the same undetermined function $I_u(x)$.  The contribution from Fig.~2(f) is then
\begin{eqnarray} 
M \tilde e (x) \biggr\vert_{\ref{BE}f + c.c }   =     -  m_q\frac{\alpha_s}{2\pi}  C_F \theta (x) \biggr [  \biggr (-\frac{2}{\epsilon_c} + \ln \frac{ \mu^2 e^\gamma}{4\pi \mu_c^2} \biggr ) \biggr ( 1 +  \frac{1+x}{1-x}  \biggr )   
  +  [ I_u (x) ]  \biggr ]  +  
 {\mathcal O} (\alpha_s m_q^3).
\end{eqnarray} 
Summing each contribution, we have the real correction of $\tilde e (x)$:
\begin{eqnarray} 
    M \tilde e  (x) \biggr \vert_R    &=& - m_q \frac{\alpha_s C_F}{2\pi} \theta (1-x) \theta (x)  \biggr (-\frac{2}{\epsilon_c} + \ln \frac{ \mu^2 e^\gamma}{4\pi \mu_c^2} \biggr )  \biggr ( 1 + \frac{1+x}{1-x} \biggr )
   +  m_q C_F \frac{\alpha_s}{2\pi} \theta (x)  [ I_u (x) ] .  
 \label{40}       
 \end{eqnarray}      
As discussed after Eq.~(\ref{33}), the function $I_u (x)$ is zero for  $x\neq 0$ and can be nonzero at $x=0$. 
With this one can easily find the sum of the real correction to $e(x)$ and the real correction to $\tilde e(x)$ is zero. We can conclude that the difference $\Delta e(x)$ is zero at one-loop level.  

\par 
One can show that $\Delta e(x)$ is zero at any order; that is, there is no difference whether we define $e(x)$ with a $T$-ordered product or without $T$-ordering.  If we use the light-cone quantization, then $x^+$ is taken as the time and QCD is then canonically quantized 
in the light-cone gauge $n\cdot G=G^+=0$. One then has the equal-time anticommutation relation for the plus-components of $\psi$ according to \cite{KS} 
\begin{equation} 
  \biggr \{ \psi ^{(+)}  (x), \bar \psi^{(+)} (0) \biggr \}  \biggr \vert_{x^+ =0} = \frac{1}{2} \gamma^- \delta (x^-) 
    \delta^2 (x_\perp) . 
\end{equation}  
As discussed before, $e(x)$ is defined as the product of one plus-component and one minus-component of $\psi$. 
The difference in this case is determined by the anticommutator of one plus-component and one minus-component. 
The anticommutator by our expressing the minus-component with the plus-component with Eq.~(\ref{17}) is 
\begin{eqnarray} 
 {\rm Tr}  \biggr \{ \psi ^{(-)}  (0 ), \bar \psi^{(+)} (\lambda n ) \biggr \}   &=&  \frac{1}{2} {\rm Tr}  \int_0^\infty d\lambda_1  \gamma^+ (\gamma_\perp \cdot D_\perp + i m_q ) \biggr \{ \psi ^{(+)}  (\lambda_1 n  ), \bar \psi^{(+)} (\lambda n ) \biggr \} 
 \nonumber\\
 &=& \frac{1}{4} \theta (\lambda ) {\rm Tr} \biggr [    \gamma^+ (\gamma_\perp\cdot  D_\perp + i m_q )\gamma^- \biggr ] 
    \delta^2 (0_\perp)   = i m_q N_c \delta^2 (0_\perp).  
\end{eqnarray}  
The anticommutator is a constant; no fields are involved. In the calculation of $e(x)$ with only connected diagrams considered, the matrix element of the constant anticommutator is excluded. 
 Hence, one can conclude that there is no difference 
whether we define $e(x)$ with a time-ordered product or an ordered product of operators. This agrees with our one-loop result above.  Since the difference is zero, the unknown function $I_u(x)$ in Eq.~(\ref{40}) in the calculation with 
cut diagrams is just  the $\delta(x)$-term in Eq.~(\ref{25}) calculated with the uncut diagram.

\par
From our detailed calculation, it is clear that the origin of the $\delta(x)$-term is due to the existence of a product of two denominators of the same quark propagator.  
 If we can use massless quark propagators instead of massive quark propagators with $q$ in Fig.~1(a), then the term does not appear, because one of two factors $q^2-m_q^2$ in the denominator in Eq.~(\ref{Fig1a}) is canceled.  But this is inconsistent with a single quark state with a nonzero mass. If we take a massless quark state, 
$e(x)$ is zero because of helicity conservation of QCD. 
 However, in reality a single quark does not exist. We  observe only hadrons. A hadron consists  of quarks, antiquarks, and gluons (i.e., it is 
 a multiparton state). If we calculate $e(x)$ of  a multiparton state, in which quarks are massless, 
 then the contribution proportional to $\delta(x)$ is absent.   The sum rule in Eq.~(\ref{SRE}) should  be satisfied without such a contribution. We examine this in the next section.    
\par 
\par
\begin{figure}[hbt]
	\begin{center}
		\includegraphics[width=6cm]{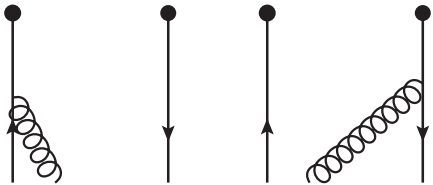}
	\end{center}
	\caption{Feynman diagrams for $e(x)$ of the multiparton state at tree level.}
\end{figure}

\par
\par\vskip5pt
\noindent 
{\bf 4. $e(x)$ of multiparton states} 
\par\vskip5pt 

We introduce the multiparton state as a superposition of a single quark state and a quark-gluon state:
\begin{equation}
 \vert n [\lambda ] \rangle  =  \vert q(p,\lambda) [\lambda ] \rangle + c_1
                   \vert q(p_1,\lambda_q) g(p_2,\lambda_g ) [\lambda ] \rangle,
\label{MPS}                    
\end{equation}
where $p_1+p_2 =p$.  The state has helicity $\lambda$. In the first term, the quark helicity is given by $\lambda_q =\lambda$.
For the quark-gluon state, the total helicity is the sum $\lambda_q + \lambda_g =\lambda$. The helicity of the single quark is always opposite that of the quark in the quark-gluon state (i.e., $\lambda_q =-\lambda$). 
The quark-gluon state is in the fundamental representation.  The quark $q$ in the multiparton state is massless. $c_1$ is a coefficient with the dimension as a mass.  All partons in the state move in the $z$-direction with $p_1^ + = x_0 p^+$ and $p_2^+ = (1-x_0) p^+=\bar x_0 p^+$. Such a multiparton state was used to study factorization problems of single transverse-spin asymmetry in \cite{MaWang,MSMP} and evolutions of chirality-odd operators in \cite{MWZ}.   
If we calculate the distribution $e(x)$ of the state, the contribution of the single massless quark state is zero because of helicity conservation of QCD. Only the interference of the single quark state with the 
quark-gluon state gives a nonzero contribution because  the helicity of the quark in the single quark state is not the same as that in the quark-gluon state.  At tree level, the contributions are from the diagrams in Fig.~3.  We have
\begin{equation}
  M e^{(0)} (x) = -  c_1 g_s C_F \sqrt{2 x_0} \biggr [ \delta (1-x) -\frac{1}{x_0} \delta (x-x_0) \biggr ].   
\end{equation}
At tree level, the matrix element is
\begin{equation} 
     \langle n \vert \bar \psi \psi \vert n\rangle^{(0)}  =  c_1 g_s C_F \sqrt{2 x_0} \frac{\bar x_0} {x_0}. 
 \end{equation} 
 At this order the sum rule is satisfied. 
 
 \par 
 At one-loop level, there are corrections from diagrams by addition of extra one-gluon exchange in the diagrams in Fig.~3. 
The calculation is straightforward. We skip the details of the calculation and give the result: 
\begin{eqnarray} 
M e^{(1)} (x) &=&  - c_1 g_s \sqrt{2 x_0}\frac{\alpha_s C_F}{2\pi}\biggr (-\frac{2}{\epsilon_c} + \ln \frac{ \mu^2 e^\gamma}{4\pi \mu_c^2} \biggr )  \biggr \{  
 -\frac{1}{2 N_c} \theta (-x) \theta (\bar x_0 +x) \frac{ \bar x_0 + x}{\bar x_0 (x_0-x ) } 
\nonumber\\
&& +\theta (x)\theta(1-x) \biggr [   C_F  \biggr ( \frac{1}{2} \biggr ( \delta (1-x) -\frac{1}{x_0} \delta (x_0-x) \biggr ) + \frac{2-x}{(1-x)_+}  + \theta (x_0-x)\frac{2 x_0 - x}{x_0^2 (x-x_0)_+ } \biggr )
\nonumber\\
   && +\frac{N_c}{4} \biggr (  \frac{2}{(1-x)_+} -  \frac{2}{x_0 (x_0-x)_+}  \theta (x_0-x) 
     \biggr ) + \frac{1}{2 N_c} \biggr ( \frac{\ln x_0}{\bar x_0} \biggr ( \delta (1-x) - \delta (x_0-x) \biggr )\nonumber\\
   &&   - \frac{x-\bar x_0}{(1-x)_+} \biggr ( \frac{1}{x_0} \theta (x-\bar x_0) 
 - \frac{1}{\bar x_0} \theta (\bar x_0 -x) \biggr ) 
+ \theta (x_0-x) \frac{\bar x_0 +x }{x_0(x_0-x)_+} \biggr ) \biggr ] 
    \biggr \},
 \label{E1L}          
\end{eqnarray}  
where  the $+$-distributions are always understood as the replacements in the integration over $x$ with 
a test function $t(x)$, 
\begin{equation} 
   \frac{ t(x) } {(1-x)_+} \to \frac{ t(x)- t(1) }{1-x}, \quad \frac{ t(x)}{(x-x_0)_+} \to \frac{ t(x)- t(x_0 )}{x-x_0}. 
\end{equation}  
The result for $e^{(1)} (x)$ in Eq.~(\ref{E1L}) can also be obtained by calculating $E_D(x_1,x_2)$ or 
$\tilde T_F (x_1,x_2)$ in the first step, and then using the relation in Eq.~(\ref{E(x)}). This will provide an important check. The result for $E_D(x_1,x_2)$ of our multiparton state at one-loop level can be extracted from the study of evolutions 
of chirality-odd twist-3 operators in \cite{MWZ}.  Agreement is found between the result in Eq.~(\ref{E1L}) and that obtained from 
$E_D(x_1,x_2)$. However, the expression for $E_D(x_1,x_2)$ is too long to be given here.

As expected, the one-loop correction in Eq.~(\ref{E1L})  has no contribution proportional to $\delta(x)$.   From $e^{(1)}(x)$ we have its first moment:
\begin{equation} 
    \int d x M e^{(1)} (x)  =  \frac{3}{4} c_1 g_s \frac{\alpha_s C_F^2}{\pi}\biggr (-\frac{2}{\epsilon_c} + \ln \frac{ \mu^2 e^\gamma}{4\pi \mu_c^2} \biggr ) \sqrt{2 x_0} \frac{\bar x_0}{x_0}. 
\end{equation}
Calculating the one-loop correction of the matrix element directly, one finds that it is exactly the above result. 
Therefore,  the sum rule in Eq.~(\ref{SRE}) is satisfied at one-loop level with our multiparton state, and 
$e(x)$ does not have a contribution proportional to $\delta (x)$.  Our result for $e(x)$ also satisfies the sum rule with the second moment. The second moment is zero in our case, because the quark is massless.  

Since a proton is, in general, a superposition of multiparton states,  
the distribution $e(x)$ of a proton has contributions not only from a single quark state but also  from interference between different states, such as a single quark state and a multiparton state. 
The contribution from a single quark state will be proportional to the quark mass from arguments from perturbation theory, while the contribution from interference will survive in the massless limit, as shown from our study here.  The contribution from interference to $M e(x)$ is expected at the order 
$1/R$, where $R$ is the size of a proton from the argument of dimension. 
This enables us to decompose $e(x)$ into two parts: 
\begin{equation} 
     M e(x) =  \frac{1}{R} e_{I} (x) +  m_q e_s (x), 
\end{equation} 
 where $e_I$ denotes the interference contribution and $e_s$  denotes the single quark contribution. 
Therefore, relative to the  interference contribution, the single quark contribution is suppressed by $m_q/m_N$, where $m_N$ is the nucleon mass. 
From combination of the result in \cite{BuKo} and our result,  the possible $\delta (x)$-term exists only in $e_s$. Therefore, 
its effect is suppressed by $m_q R \sim m_q/m_N$.  If we ignore the quark mass, it is expected that $e(x)$ will not contain a contribution proportional to $\delta(x)$.  It should be kept in mind that the conclusion drawn here 
is based on arguments from perturbative QCD.

\par 
From \cite{BuKo}  the contribution with $\delta(x)$ exists not only in $e(x)$ but also in $h_L(x)$, a twist-3 distribution of a longitudinally polarized proton. Our results for $e(x)$ do not apply for the case 
of $h_L$.  The reason is as follows: 
For a single quark state $h_L(x)$ also has a $\delta(x)$-contribution from Fig.~1(a) as 
$e(x)$ does. As mentioned at the end of Sect.~3. if we take massless quark propagators in Fig.~1(a), 
$e(x)$ does not have a $\delta(x)$-contribution. But $h_L(x)$ still has such a contribution if the quark propagators
are massless.   
It seems that  the existence of such a contribution in higher-twist parton distributions is quite general.  The origin of a $\delta(x)$-contribution may be zero modes of partons and their long-range order inside hadrons, as discussed in \cite{XJi}.  The zero modes can result in the quark fields at infinity of space-time 
being nonzero. In this case, the solution in Eq.~(\ref{17}) should be modified as
\begin{eqnarray} 
   \psi^{(-)}(x) = \psi^{(-)}(x)\biggr\vert_{x^- =\infty} + \frac{1}{2} {\mathcal L}_n^\dagger (x)  \int_0^{\infty} d \lambda \biggr [ {\mathcal L}_n \gamma^+ \biggr ( \gamma_\perp^\mu  D_\mu + i m_q \biggr )  \psi^{(+)} \biggr ] ( \lambda n +x).
\label{17N}    
\end{eqnarray}
Then the sum of the three $\delta (x)$-terms in Eq.~(\ref{ER14}) is not zero. A contribution proportional 
to $\delta (x)$ can exist as
\begin{equation} 
M e(x) =\frac{1}{2} \delta (x) \langle P  \vert  \bar\psi^{(+)}(0) \psi^{(-)} (\infty n)+  \bar\psi^{(-)}(0) \psi^{(+)} (-\infty n)  \vert P\rangle + \cdots . 
\end{equation} 
However, it is unclear how the $\delta (x)$-contribution in Sect.~3 from perturbation theory is related to the zero-mode contribution, because the contribution is a nonperturbative quantity. It can be studied only with 
nonperturbative methods.  There is evidence of a $\delta(x)$-contribution  from the study of the distribution in chiral quark soliton models \cite{CQSM}.  It is possible to study it  with large-momentum effective field theory \cite{XJiLa,LaMet} through lattice QCD simulations as argued in \cite{XJi}.  

\par
\par\vskip5pt
\noindent 
{\bf 5. Summary} 
\par\vskip5pt  

We have shown that at the operator level one cannot find a contribution proportional to $e(x)$ 
if quark fields at infinity of space-time are zero.  It is true that $e(x)$ of a single quark has such a contribution in perturbation theory. However,  $e(x)$ of a multiparton state containing massless quarks has no  such contribution, as shown through our one-loop calculation. 
Since a hadron is a superposition of multiparton states, we can decompose $e(x)$ of a proton as the sum of the contribution from a single quark and that from interference of different states. 
On the basis of arguments from perturbative QCD, the single quark contribution can have a contribution with $\delta(x)$ but proportional to the quark mass, and the interference contribution is nonzero with massless quarks and has no contribution proportional to $\delta (x)$. Therefore, 
it is expected that the effect of the contribution with $\delta(x)$ is suppressed by the quark mass. If we ignore the masses of light quarks in a proton, the sum rule of $e(x)$ related to the pion-nucleon $\sigma$-term 
is not violated. In this case, one can still use the sum rule to determine the $\sigma$-term with arguments 
from perturbation theory.

\par\vskip40pt
\noindent
{\bf Acknowledgments}
\par
The work was supported by the National Natural
Science Foundation of China (no. 11675241, no. 11821505, no. 11947302 and no. 12065024) and the Strategic Priority Research Program of the Chinese Academy of Sciences (grant no. XDB34000000).

\par

\end{document}